# Smooth Curve from noisy 2-Dimensional Dataset

Avik Kumar Mahata[1], Utpal Borah[2], , Aravind Da Vinci[3], B.Ravishankar[4,] Shaju Albert[5]

[1,4]Material Science and Engineering, National Institute of Technology, Trichy, India

[2,3]Metal Forming & Tribology Programme, Materials Technology Division, Indira Gandhi Center for Atomic Research, Kalpakkam, Tamilnadu, India

[5]Materials Technology Division, Indira Gandhi Center for Atomic Research, Kalpakkam, Tamilnadu, India

**Abstract:** In this paper we will be represent the transformation of a noisy dataset into a regular and smooth curve. We performed torsion test on 15-Cr 15-Ni Titanium modified austenitic stainless steel up to its rupture. We did it these torsion tests multiple strain rate varying from 0.001-25/S and obtained huge number of data points has been obtained from with a data acquisition rate of 100 Hz. We also have few more data of 1500Hz of the same experiment on a different material (316 L Austenitic Stainless Steel). Machine is actually acquiring only torque value and the angle of rotation. The torque vs. twist data itself will be having a noise, which gets multiplied when we take the first derivative of torque-twist data. The noise becomes huge and it fluctuates from desired material properties, although some serrated flow should be present but the range of serration should not be as the fluctuation observed in our curve so smoothing was necessary. We will be documenting final shear stress-strain curve we achieved through Nonparametric Regression (Lowess and Loess), Savitzky–Golay filtering, and also the robust Nonparametric Regressions, body splines and shape preserving curves. The most acceptable one is nonparametric regression method as it filter the data without presuming the shape of the curve, as for this reason the higher order polynomials fails predicting the shape of our desired shear stress strain curve. We didn't get expected curve using splines. In this analysis we had to take further precaution as the shear stress strain curve for fully rupture test in torsion for titanium modified stainless steel (most probably none of the structural material like steel, aluminum, copper etc. has full rupture curve, fatigue test can be found in literature) was not available before. There will be few discussions on the experiment also but the main aim of this discussion is the statistical data analysis.

1.  Introduction:

Smoothing a dataset consists in finding approximate solutions of a function that captures important patterns in the data at each data point, while disregarding noise or other fine-scale structures. Smoothing of data is a necessary step for any laboratory experiment in any fields of study. Progress in modern day computers and computational tools (Matlab, Mathematica etc.) gives an ease to data analysis at a huge scale. Currently we in data mining

2.  Experiment

Experiment was done on a cylindrical specimen size of 10mm diameter and 25mm length with a total length of 120mm, 30mm in both side used for holding. Titanium modified austenitic stainless steel (also known as alloy D9) is used as a cladding and wrapping material in fuel assembly in a Prototype Fast Breeder Nuclear Reactor. Twist was applied on the specimens up to its failure. There are a lot of mechanical and metallurgical can be done on the nature of failure of titanium modified austenitic stainless steel but here we will limit our discussion on the nature of the curves obtained from the shear stress-strain data. The curves are showing desirable properties of the material like hardening, softening, serrated flow etc. Following are the stress strain plots obtained from our experiment without any modification. We only neglected the negative values as the negative values don't have any physical significance now for data analysis. From mechanics point of view negative stress is the resistance of materials against the deformation.

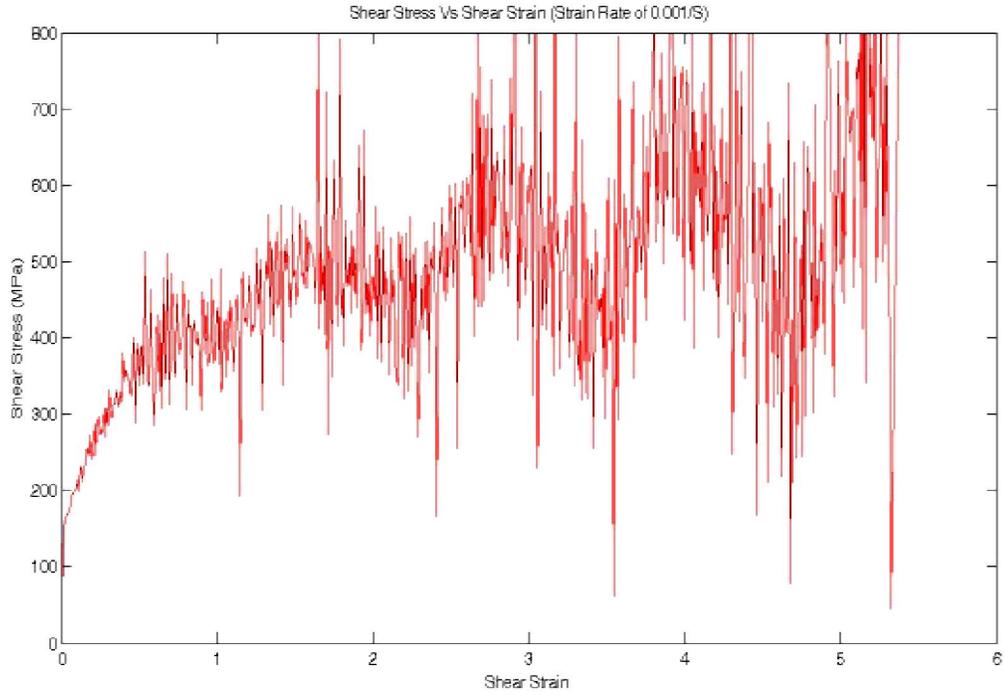

Fig1. The original plot at 0.001/S Strain Rate where we recorded around 0.3 million data points

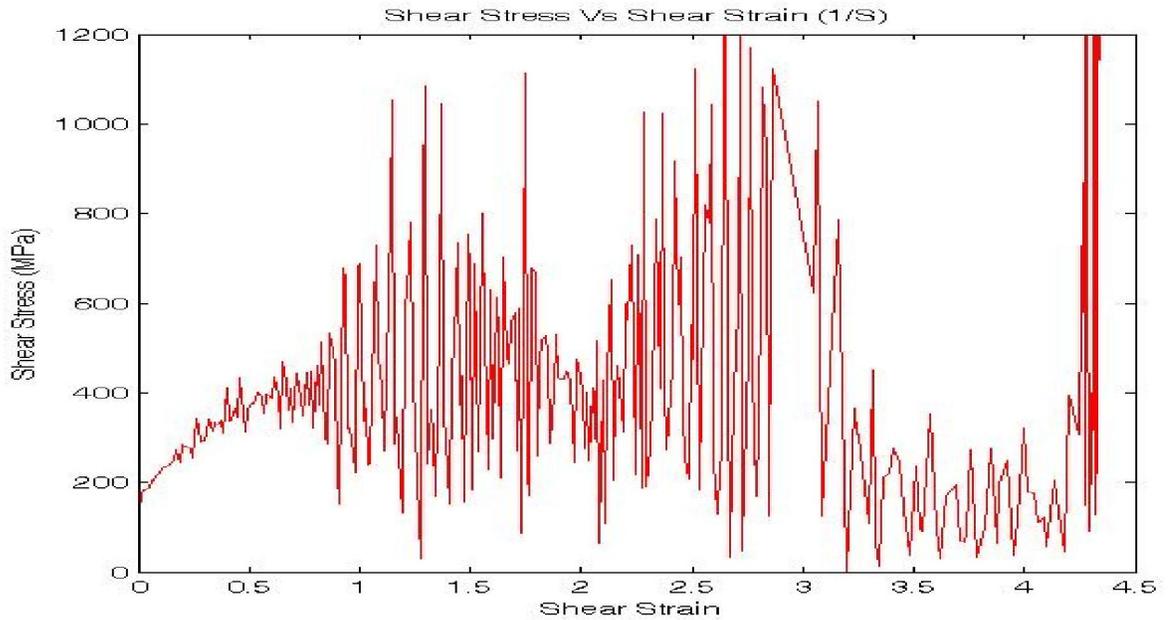

Fig2. The original plot at 1/S Strain Rate where we recorded around 10500 data points

We recorded our data at 100 Hz , when the experiment was being done on 0.001/S strain rate it recoded for a longer period than 1/S, that's why the noise is also very high in lower strain rate. We need a careful observation of the plot of 0.001/S strain rate, which will be trend of the successive stain rate plots.

## 3. Review on Smoothing Techniques

### 3.1 Nonparametric Regression

*Nonparametric simple regression* is called scatter plot smoothing, because the method passes a smooth curve through the points in a scatter plot of *y* against *x*. Scatter plots are omnipresent in statistical data analysis and presentation. Nonparametric simple regression forms the basis, by extension, for nonparametric multiple regression, and directly supplies the building blocks for a particular kind of nonparametric multiple regression called *additive regression*.

**Binning** suppose the predictor variable *x* is discrete, we want to the average value of *y* (or some other characteristic of *y*) which changes with *x*; now we can dissect the range of *x* into a large number of narrow class intervals called *bins*. Because each bin is narrow, these bins averages do a good job of estimating the regression function anywhere in the bin, including it's center. But this binning will mostly work for large sample, smaller sample is not practical to bin.

- There will be few observations in each bin, making the sample bin averages unstable
- To calculate stable averages, we need to use a relatively small number of wider bins, producing cruder estimate of the population regression function.

There are two obvious ways to proceed:

*i)* We could dissect the range of *x* into bins of equal width. This option is attractive only if *x* is sufficiently uniformly distributed to produce stable bin averages based on a sufficiently large number of observations.

*ii)* We could dissect the range of *x* into bins containing roughly equal numbers of observations

**Local Averaging** The essential idea behind local averaging is that, as long as the regression function is smooth, observations with *x*-values near a focal $x_0$ are informative about $f(x_0)$.

- Local averaging is very much like binning, except that rather than dissecting the data into non-overlapping bins, we move a bin (called a window) continuously over the data, averaging the observations that fall in the window.
- We can calculate *f(x)* at a number of focal values of *x*, usually equally spread within the range of observed *x*-values, or at the (ordered) observations, $x_{(1)}, x_{(2)}, \ldots, x_{(n)}$.
- As in binning, we can employ a window of fixed width *w* centered on the focal value $x_0$, or can adjust the width of the window to include a constant number of observations, *m*. These are the *m* nearest neighbors of the focal value.
- Problems occur near the extremes of the *x*'s. For example, all of the nearest neighbors of $x_{(1)}$ are greater than or equal to $x_{(1)}$, and the nearest neighbors of $x_{(2)}$ are almost surely the same as those of $x_{(1)}$, producing an artificial flattening of the regression curve at the extreme left, called boundary bias. A similar flattening occurs at the extreme right, near $x_{(n)}$.
- In addition to the obvious flattening of the regression curve at the left and right, local averages can be rough, because *f(x)* tends to take small jumps as observations enter and exit the window. The kernel estimator (described later) produces a smoother result.

- Local averages are also subject to distortion when outliers fall in the window, a problem addressed by robust estimation.

**Linear Model The** Regression analysis comes in different flavors. In the classical linear model, the symbol Y denotes the dependent variable such that its conditional expectation is a linear function of the observed value x of the explanatory variable or its arbitrary transformation, i.e.,

$$E(Y|X = x) = f(x)^T \beta; \qquad (1)$$

Where $f(x) = (f_1(x), \ldots, f_p(x))^T$ is a known function of $x$, β is a p-dimensional vector of unknown parameters, and $Var(Y|X = x) = \sigma^2 > 0$. Often, we introduce an artificial random variable ξ, the so called random error, in order to write the linear model in the most common way:

$$Y = f(x)^T \beta + \xi, \qquad (2)$$

Where the unobservable random error ξ is centered and has variance $\sigma^2$, i.e., Eξ = 0 and $E\xi^2 = \sigma^2$.

In the above linear model, we assume that the functional form of dependency is known in advance and it remains to estimate only the vector of unknown parameters β. For example, by setting $f(x) = (1, x)^T$ we obtain a straight line with intercept given by $\beta_0$ and slop equal to $\beta_1$. Parabolic and cube function may be obtained by choosing $f(x) = (1, x, x^2)$ or $(x) = (1, x, x^2, x^3)^T$, respectively.

An estimator β' of the unknown parameter β is usually calculated from $n$ observations of pairs $(Y_i, x_i)$ satisfying (2), i.e.,

$$Y_i = f(x_i)^T \beta + \xi_i, \ i = 1, 2, \ldots, n \qquad (3)$$

Assuming that the random errors $\xi_i$, $i = 1,2,\ldots,n$, are independent and identically distributed. The n equation (3) are often rewritten in Matrix notation:

$$y = x_F \beta + \xi, \qquad (4)$$

Where $y = (Y_1, Y_2, \ldots Y_n)^T$ is the response vector, $x_F$ is called *design matrix* with rows $f(x_i)$, and $\xi = (\xi_1, \ldots \xi_2)^T$ is a centered random vector with variance matrix $\sigma^2 I_n$.

Assuming the design matrix $x_F$ has full rank, equation (4) allows to express the Least Squares (LS) estimator of β as:

$$\beta'^{LS} = (x_F^T x_F)^{-1} x_F^T y, \qquad (5)$$

**The Variance of Least Squares Estimator.** In order to construct confidence interval for components, one needs an estimator of covariance matrix $\beta^{,LS}$. Now, it can be shown that, given x, the covariance matrix of the estimator $\beta^{,LS}$ is equal to,

$$(x_F^T x_F)^{-1} \sigma^2,$$

Where $\sigma^2$ is the variance of the noise. We can get the measure of residuals from variance. Which is,

$$e_i = y_i - x_{i,1} \beta_1'^{LS} - \cdots x_{i,p} \beta_p'^{LS} \qquad (6)$$

The covariance matrix $\beta'^{LS}$ can, therefor be estimated by

$$(x_F^T x_F)^{-1} \sigma'^2, \text{ where,}$$

$$\sigma'^2 = \frac{i}{n-p}\sum_{i=1}^{n} e_i^2 \tag{7}$$

We need to remove the parametric assumption from (2) to obtain *non-parametric regression model*:

$$Y = E(Y|X = x) + \xi = m(x) + \xi \tag{8}$$

Where the symbols carry the meaning mentioned before. An example of non parametric regression estimator is shown below.

Compared to the linear model (2), the nonparametric regression model (8) is more flexible. On the other hand, the nonparametric regression estimator is not as easily interpretable and it is often used only as a graphical tool. Additionally, one also has to choose an appropriate value of some smoothing parameters that typically control the smoothness of the estimator.

An overview of nonparametric regression (or smoothing) methods may be found, e.g., in Hardle (1990); Simono (1996); Fan and Gijbels (1996); Hardle et al. (2004). The standard smoothing approaches include splines, wavelets, moving averages, running medians, local polynomials, regression trees, neural networks, and other methods. From now on,

We concentrate on the kernel method: the kernel regression estimators are defined as locally weighted averages and its properties may be derived quite easily.

### 3.1.1 Kernel Regression

In this section, we introduce to basic notation and provide a short review of our results concerning the kernel regression estimator based on the nonparametric regression model (8).

In practice, it is important to distinguish fixed and random design experiments. In a fixed design experiment, we choose the values of the explanatory variable according to a certain rule, e.g., as a qualities of certain probability distribution. In a random design experiment, we may control only the probability distribution of the explanatory variable but the observed values are random.

**Random Design** We assume that model (8) holds and that we observe pairs of random variables $(X_i, Y_i)$, $i=1,...,n$ such that:

$$Y_i = m(X_i) + \xi_i \tag{9}$$

Where $E\xi_i = 0$ and $Var\xi_i = \sigma^2$. The unknown regression function *m* may be estimated, for example by using classical ***Nadaraya-Watson estimator*** *(Nadaraya; 1964; Watson; 1964)*:

$$w_b^{NW}(x) = \frac{\sum_{i=1}^{n} K_b(x-X_i)Y_i}{\sum_{j=1}^{n} K_b(x-X_j)} \tag{10}$$

Where $K_b(x) = K(x/b,)$ *K* is a *kernel function* and b>0 is a *bandwidth*.

**Common examples of kernels**

• Gaussian density,

• Epanechnikov: $K_b = \frac{3}{4}(1 - t^2)$, $t^2$<1,0 otherwise.

• Minimum var: $K_b = \frac{3}{8}(3 - 5t^2)$, $t^2$<1,0 otherwise.

**3.1.2 Loess/Lowess** Kernel still exhibits bias at the end points. For that we should use kernel weights to estimate a running line. One way of looking at scatter diagram smoothing is as a way of depicting the "local" relationship between a response variable and a predictor variable over parts of their ranges, which may differ from a "global" relationship, determined using the whole data set. Loess is a bivariate smoother function or procedure to produce a smooth curve through scatter diagram. Curve is drawn in such a way that it will be having some desirable properties. In general, the properties are that the curve indeed be smooth, and that locally, the curve minimize the variance of the residuals or prediction error. The bivariate smoother used most frequently in practice is known as a "lowess" or "loess" curve. The method consequently makes no assumptions about the form of the relationship, and allows the form to be discovered using the data itself. (The difference between the two acronyms or names is mostly superficial, but there is an actual difference in R–there are two different functions, lowess and loess, which will be explained below.)

'Lowess' and 'loess' are two different functions in R which implement local regression. 'loess' is the more recent program and it is essentially equivalent to 'lowess' except that it has more features.

The R function 'loess' has three things which 'lowess' doesn't:

1. It accepts a formula specifying the model rather than the x and y matrices

2. As you've noted, it can be used with more than one predictor.

3. It accepts prior weights.

4. It will estimate the "equivalent number of parameters" implied by the fitted curve. On the other hand, 'loess' is much slower than 'lowess' and occasionally fails when 'lowess' succeeds, so both programs are kept in R. When there is only one predictor variable and no prior weights, 'lowess' and 'loess' are in principle exactly equivalent. The only aspect in which it is not possible to make 'loess' and 'lowess' agree exactly is in their treatment of large data sets. When x and y are very long, say 10s of thousands of observations, it is impractical and unnecessary to do the local regression calculation exactly, rather it is usual to interpolate between observations which are very close together. This interpolation is control by the 'delta' argument to 'lowess' and the 'cell' and 'surface' arguments to 'loess'. When there are a large number of observations, 'lowess' groups together those x-values which are closer than a certain distance apart. Although grouping observations based on distance is in-principle the best approach, this is impractical for 'loess' because 'loess' is designed to accept many x-variables. So 'loess' instead groups observations together based on the number of observations on a cell rather than distances. Because of this small difference, 'lowess' and 'loess' will almost always give slightly different numerical results for large data sets. The difference is not generally important. ([Department of Statistics, ETH Zurich](#)). In Matlab we can call both the function and also access few more estimation techniques like Savitzky-Golay Filter (A. Savitzky and M. J. E. Golay, "Soothing and differentiation of data by simplified least squares procedures," Anal. Chem., vol. 36, pp. 1627–1639, 1964; [Ronald W. Schafer, UC Berkeley](#)) .

**Obtained Plots from our data** The *shear stress shear strain* (σ-γ) can be derive from the obtained from *torque-twist (M-θ)* (Need to keep in mind that machine can only obtain the torque and twist data). Gage section geometry determines the deformation level and it's rate for a given amount of twist and twist rate (J.J Jonas, Chapter 8, Process Design and Workability handbook).

$\gamma = \frac{r\theta}{L}$    &    $\dot{\gamma} = \frac{r\dot{\theta}}{L}$    L is the length of specimen and r is the diameter.

**Reduction of Torque-Twist Data to Shear Stress versus Shear Strain** When the specimen is twisted the component of torque *dM* due to one such tube of radius and a thickness *dr*. We can deduce the formula for torque and shear stress like below like below when *dr=Ldγ/θ* ;

$$M = \frac{2\pi}{\theta^3} \int_{\gamma_1}^{\gamma_2} \tau \gamma^2 \tag{11}$$

$$\tau_a = \frac{1}{2\pi r^3}\left(3M + \theta \frac{dM}{d\theta}\right) \tag{12}$$

So the plots below will be obtained after smoothing using loess/lowess.

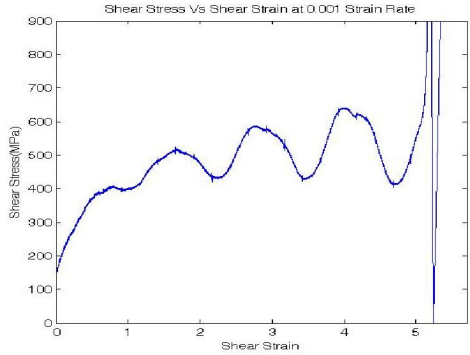 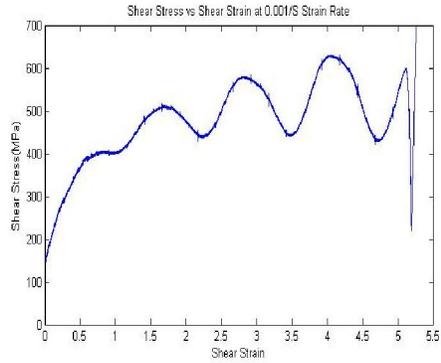

Fig:3  Loess Smoothing with a size of 0.1                Fig 4: Lowess Smoothing with a size of 0.1

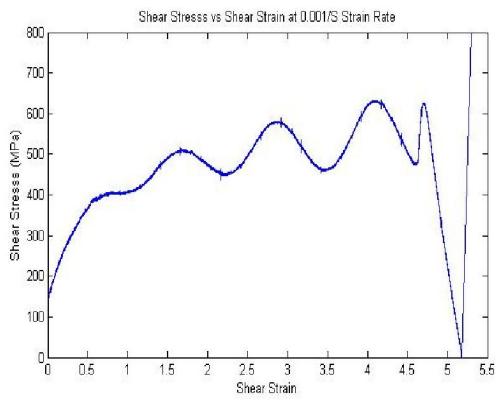 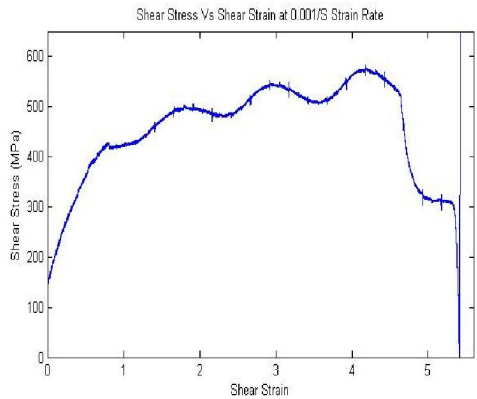

Fig5:  Loess Smoothing with a size of 0.3                Fig6: Lowess Smoothing with a size of 0.3

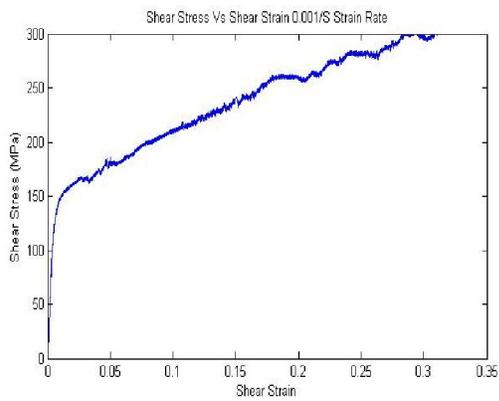 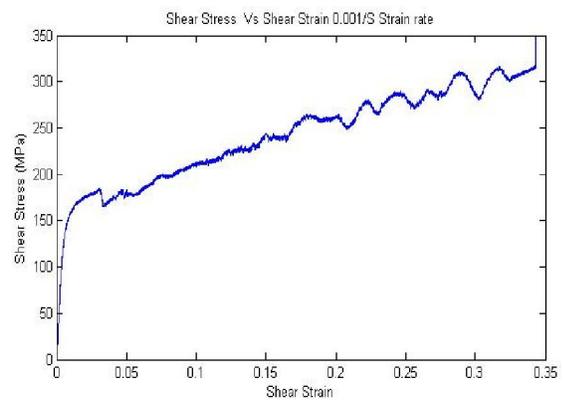

Fig7: Robust lowess for first 20000 datapoints           Fig8:Robust lowess for first 20000 datapoint

***Discussion on Loess/Lowess Smoothing*** Above all are the plots obtained after loess/lowess smoothing. There is not much differences in the plots by loess and lowess smoothing. Only the end bias can be observed in loess smoothing. But if we go up to bias size 0.3 then it differs from each other, later if we try robust lowess and loess the progression of both the plots differs. Here we have to interpret the data with our knowledge in mechanics of failure. Generally we always observe the hardening and softening pattern in a failure. We observe that in our torque-twist diagram. That hardening softening should be reflected in the final stress strain diagram. Here in Fig 3 & 4 we are observing the regular stress strain diagram. We are not keeping serrated flow in the consideration as that effect can be observed in a very small fluctuation. We can discuss it in another technical report as that will involve different areas of research which is not relevant to the current discussion. So, as of now we can consider Fig3 and 4 for analyzing the torsional behavior of alloy D9.

**3.2 Splines** A *spline* is a piecewise polynomial with pieces defined by a sequence of *knots or nodes*,

$$\xi_1 < \xi_2 < .... < \xi_k$$

such that the pieces join smoothly at the *knots/nodes*.

A *n*th degree *spline* function *f(x)* is a piecewise *n*th degree polynomial function. The polynomials are joined at the *knots* in such a way that *(n-1)*th are continuous. Within these constraints, the function *f(x)* is selected which minimizes:

$$\sum (f(x_i) - x_i)^2 + p. \int (f^{\frac{n+1}{2}}(x))^2) dt$$

Where $x_i$ is the raw data sample, and $f^k$ denotes the *k*th derivative of *f(x)*. The weight factor *p* is the smoothing parameter. With *p=0*, an interpolating spline will be obtained, with *p=∞* a least squares fit of the entire dataset using a single polynomial of degree *(n-1)/2*.

The general idea of a spline is like on each interval between data points, represent the graph with a simple function. The simplest spline is something very familiar to us; it is obtained by connecting the data with lines. Since linear is the most simple function of all, linear interpolation is the simplest form of spline, called *linear spline*. The next simplest function is quadratic. If we put a quadratic function on each interval then we should be able to make the graph a lot smoother. If we were really careful then we should be able to make the curve smooth at the data points themselves by matching up the derivatives. This can be done and the result is called a *quadratic spline*. Using cubic functions or 4th degree functions should be smoother still. There is an almost universal consensus that cubic is the optimal degree for splines and so we focus on *cubic splines* or higher order spline.

**Cubic Spline** If there are *n* data points, then the spline *S(x)* is the function:

$$S(x) = \begin{cases} C_1(x), & x_0 \leq x \leq x_1 \\ C_i(x), & x_{i-1} \leq x \leq x_i \\ C_n(x), & x_i \leq x \leq x_n \end{cases} \qquad (13)$$

Where each $C_i$ is a cubic function. The most general function has the form :

$$C_n(x) = a_i + b_i x + c_i x^2 + d_i x^3$$

To determine the spline we must determine the coefficients, $a_i, b_i, c_i$ and $d_i$ for each *i*. Since there are *n* intervals, there are *4n* coefficients to determine. First we require that spline be exact at the data:

$$C_i(x_{i-1}) = y_{i-1} \text{ and } C_i(x_i) = y_i,$$

at every data point. To make *S(x)* as smooth as possible we require:

$$C_i'(x_i) = C_{i+1}'(x_i)$$

$$C_i''(x_i) = C_{i+1}''(x_i)$$

The usual is to require:

$$C_1''(x_0) = C_n''(x_n) = 0$$

This is called *natural* or *simple* boundary conditions. The other common option is called *clamped* boundary condition:

$$C_1'(x_0) = C_n'(x_n)$$

A *natural spline* is linear outside the range of the data.

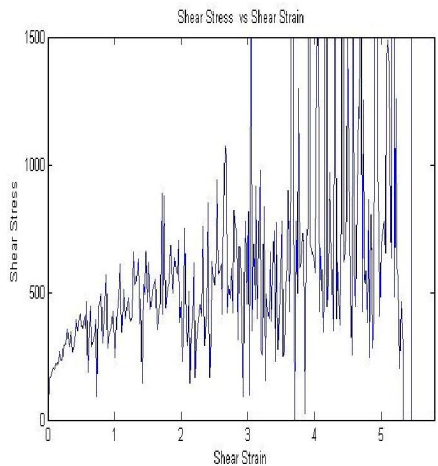
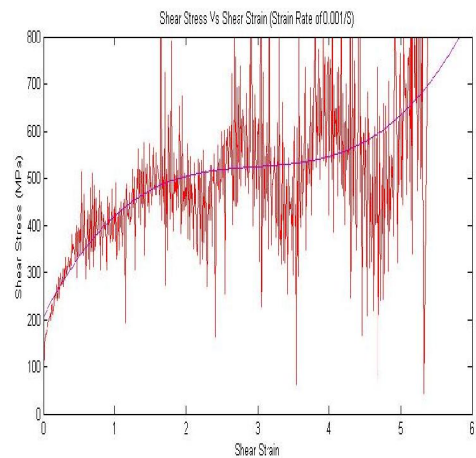

Fig9: Natural Spline Interpolant    Fig10: Cubic Spline Interpolant

***Discussion*** Spline is failing predicting the correct behavior of the curve. Cubic and natural spline are trying to interpret data into 2 different ways, and both failing for our application. Our data points will be having a local and a global softening and hardening rate. So, cubic spline completely neglecting all the local effects and giving one very local values.

**B-Splines** The power series representation is useful for understanding splines but is not well suited for computation because successive terms tend to be highly correlated. It is, however, very easy to use. A much better representation of splines for computation is as linear combinations of a set of basis splines called *B-splines*. These are numerically more stable, among other reasons because each *B-spline* is non-zero over a limited range of knots. They are not so easy to calculate, but fortunately Matlab provides curve fitting toolbox where we can compute the *B-Spline* . The main problem is where to place the knots. Often they are placed at selected quantiles (i.e. the terciles, or quartiles, or quintiles, depending on how many knots we want). A smarter strategy would place more knots in regions where f (x) is changing more rapidly.

**Hermite Polynomials** Polynomials are convenient for interpolation, we can manipulate them symbolically, we can calculate them fast, and according to Weierstrass approximation theorem any continuous function on some interval [a,b] can be approximated by polynomials. In practice polynomials having degrees higher than cubic doesn't give proper approximation. An alternative approach that retains the advantages of working with polynomials is to work with piecewise polynomial functions. In Matlab *PCHIP* is an inbuilt function, which is also called shape preserving curve, is nothing but piecewise cubic polynomials that satisfy Hermite interpolation condition.

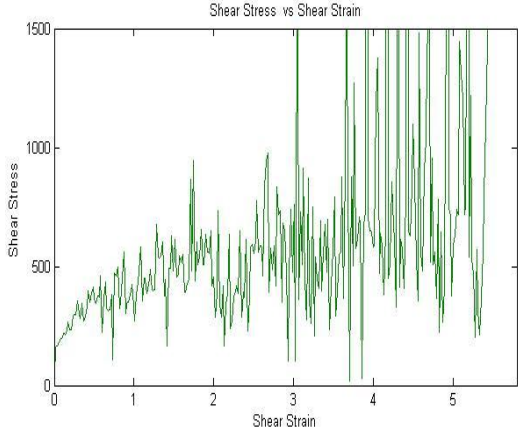

Fig11: PCHIP interpolation

**4. Outliners** An *outlier* is an observation that appears to deviate markedly from other observations in the sample. Identification of potential outliers is important. An outlier may indicate bad data. For example, the data may have been coded incorrectly or an experiment may not have been run correctly. If it can be determined that an outlying point is in fact erroneous, then the outlying value should be deleted from the analysis (or corrected if possible). In some cases, it may not be possible to determine if an outlying point is bad data. Outliers may be due to random variation or may indicate something scientifically interesting. In any event, we typically do not want to simply delete the outlying observation. However, if the data contains significant outliers, we may need to consider the use of robust statistical techniques ([NIST](NIST)).

But here in our data the fluctuation of shear stress value is extremely close and 'deleteoutliners' are not that helpful in our application. .

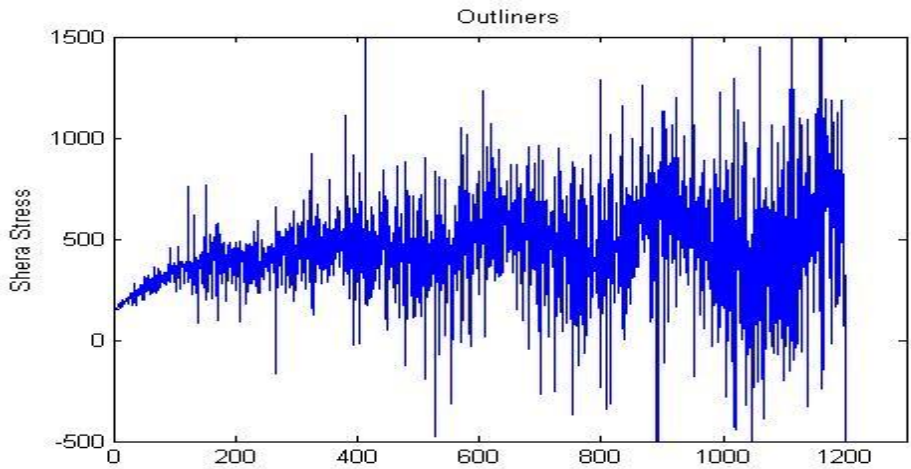

Fig12: Outliners in Fig 1 data

**5. Savitzky-Golay Filter** A much better procedure than simply averaging points is to perform a least squares fit of a small set of consecutive data points to a polynomial and take the calculated central point of the fitted polynomial curve as the new smoothed data point. In this study, a low-pass digital filter known as the *Savitzky-Golay* from (*Smoothing and Differentiation of Data by Simplified Least Squares Procedures*.; Anal. Chem., 1964, 36 (8), pp 1627–1639) was used. It is derived directly in the time domain from a particular formulation of the data smoothing problem in the time domain. Historically, this filter was used to extract the relative widths and heights of spectral lines in noisy data. It works by performing a least-squares fit of a polynomial of degree $M$, using an additional number $n_L$ of points to the left and some number $n_R$ of points to the right of each desired $x$ value. The estimated derivative is then the derivative of the resulting fitted polynomial.

The smoothing effect of the Savitzky-Golay algorithm is so aggressive that in the case of the moving average and the loss and/or distortion of vital information is comparatively limited. However, it should be stressed that both algorithms will lose some part of the original information is lost or distorted. Filter design is available in Matlab Help.

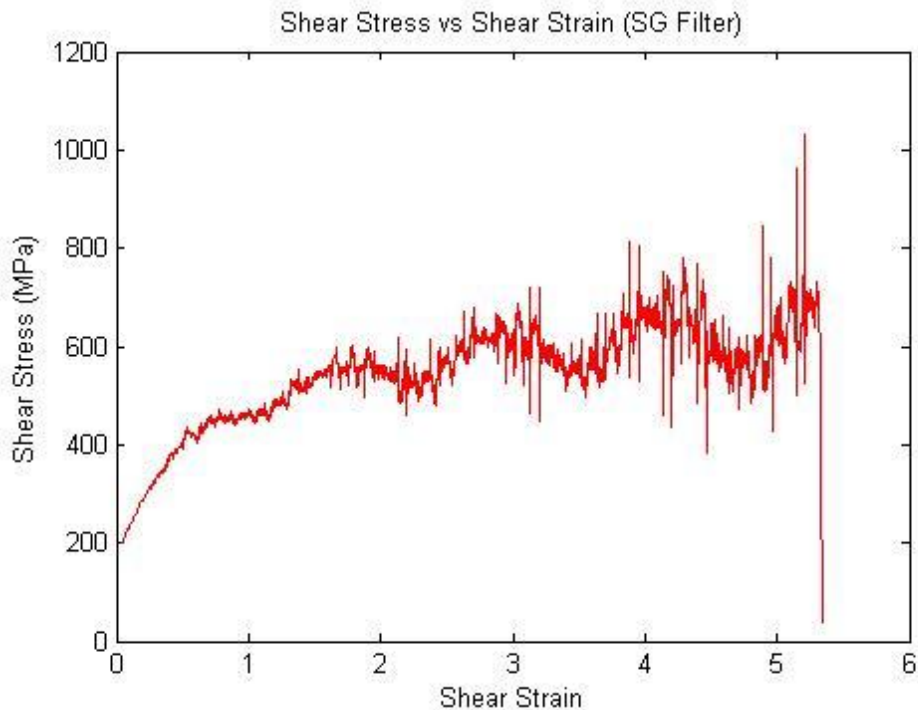

Fig 14: S-G filter for Fig 1 dataset with 3$^{rd}$ dregree polynomial & framewidth 99

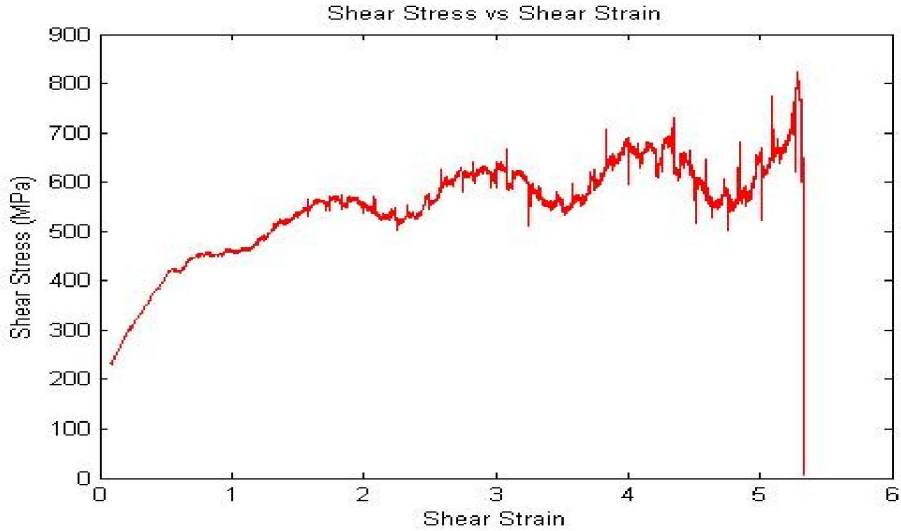

Fig 14: S-G filter for Fig 1 dataset with 3$^{rd}$ dregree polynomial & framewidth 99

We can see in the above figures that the plots are following same trend like Fig 3 & 4, but didn't cancel out all the noise. In loess/lowess we may loose a large number of data which is actually necessary to understand the material property. We can't overlook for the sake of smoothing the actual material response in a certain physical condition.

## 6. Conclusion

Choosing an optimal filter for any datasets is typically an interactive trial and error process. The goodness of a filter is best based on visual inspection of the results. Also note that it depends on the subsequent steps in the analysis too. For instance, a good filter for peak detection may be one which reduces noise to 1% of the signal. The same filter may not be good for differentiation, because the noise in the derivative is still to high. Filters for differentiation typically need a lower cut-off frequency.

***Rodney Hill*** proposed the shear stress,

$$\boldsymbol{\tau_a} = \frac{1}{2\pi r^3}\left(4M + \theta^2 \frac{d(M/\theta)}{d\theta}\right) \tag{14}$$

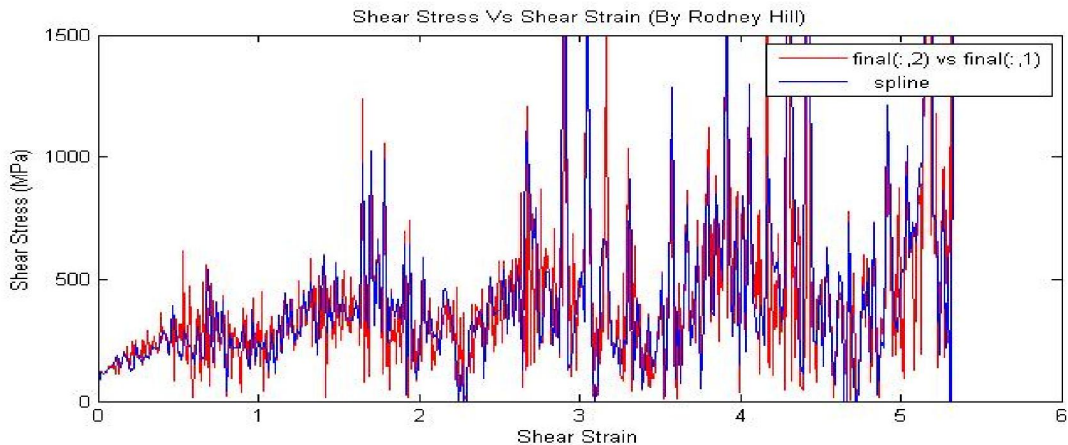

Fig 12:

But in this study also we are getting same kind of behavior of the plot. So we can conclude here choosing any of the equation we can apply *Savitzky-Golay* Filter or *loess/lowess* filter according to our requirement. But it is clear from the above discussion that we can use *Savitzky-Golay filter* for keeping maximum information from the material response.